\documentclass[12pt]{iopart}

\usepackage{iopams}

\usepackage{amssymb}
\usepackage{latexsym}
\usepackage{euscript}
\usepackage{graphics}
\usepackage[all]{xy}

\newcommand{\be}{\begin{equation}}
\newcommand{\ee}{\end{equation}}
\newcommand{\ba}{\begin{eqnarray}}
\newcommand{\ea}{\end{eqnarray}}
\newcommand{\baa}{\begin{eqnarray*}}
\newcommand{\eaa}{\end{eqnarray*}}
\newcommand{\bb}{\begin{thebibliography}}
\newcommand{\eb}{\end{thebibliography}}

\renewcommand{\bi}[1]{\bibitem{#1}}
\newcommand{\lab}[1]{\label{#1}}
\newcommand{\re}[1]{(\ref{#1})}

\eqnobysec

\renewcommand\t{\tilde}

\begin{document}
\title[Dual -1 Hahn polynomials and perfect state transfer]{Dual -1 Hahn polynomials and perfect state transfer}
\author{Luc Vinet $^1$ and Alexei Zhedanov$^2$}
\address{$^1$ Centre de recherches math\'ematiques
Universit\'e de Montr\'eal, P.O. Box 6128, Centre-ville Station,
Montr\'eal (Qu\'ebec), H3C 3J7}
\address{$^2$ Institute for Physics and Technology,
R.Luxemburg str. 72, 83114 Donetsk, Ukraine}

\begin{abstract}
We find all the $XX$ spin chains with perfect state transfer (PST)
that are connected with the dual -1 Hahn polynomials $R_n(x;
\alpha,\beta,N)$. For $N$ odd we recover a model that had already
been identified while for $N$ even, we obtain a new system
exhibiting PST.
\end{abstract}



\maketitle

\section{Introduction}
\setcounter{equation}{0} The transfer of a quantum state is said
to be perfect if there is probability 1 of finding after some time
at an end point the state introduced as input  at an initial site.
It has been realized that perfect state transfer (PST) can be
achieved in inhomogeneous $XX$ spin chains in particular, provided
the 1 -excitation energy eigenvalues satisfy a simple spacing
condition. (For reviews, see \cite{Albanese}, \cite{Kay}.) It has
further been shown recently \cite{VZ_PST} that a unique (up to
trivial rescaling) $XX$ spin chain with nearest neighbor
interactions and with PST, corresponds to each such spectrum. This
matching proceeds through the association of families of
orthogonal polynomials in a discrete variable to $XX$ spin chains
with PST. The 1-excitation eigenvalues thus intervene as the
orthogonality grid points. We have provided an algorithm to
construct the $XX$ Hamiltonians from the spectral data and have
shown how to obtain different models with PST from a parent system
with that property by performing spectral surgery \cite{VZ_PST}.

The exact solvability of the PST models is intimately  related to
the characterization of the associated orthogonal polynomials.
Among the Dunkl or -1 orthogonal polynomials that we have
discovered lately \cite{BI}, one family, namely that of the dual
-1 Hahn polynomials, interestingly relates to PST. We shall here
obtain the spin chains with PST to which they are associated and
find in the process, a new exactly solvable model.

The outline of the paper is the follows. The dual -1 Hahn
polynomials and their relevant properties are recalled in section
1. The relation between perfect state transfer in $XX$ chains and
orthogonal polynomials theory is reviewed in section 2. Finally,
depending on the parity of $N$-the number of sites minus one  -
two spin chains with PST are identified and analyzed in section 3.
The one for $N$ odd had already been reported in \cite{Shi} and
further studied in connection with dual Hahn polynomials in
\cite{SJ}. The one for even $N$ had escaped attention so far. We
conclude the paper by showing how these two models are related.

\section{Dual -1 Hahn polynomials}
\setcounter{equation}{0} Dual -1 Hahn polynomials $R_n(x;\alpha,
\beta,N)$ were introduced in \cite{-1_Hahn} as $q=-1$ limits of
the dual $q$-Hahn polynomials.

These polynomials  satisfy the 3-term recurrence relation \be
R_{n+1}(x) + b_n R_n(x) + u_n R_{n-1}(x) = xR_n(x) \lab{rec_R} \ee
and depend on an integer $N=1,2,\dots$ and two real parameters
$\alpha,\beta$.

The expression of the recurrence coefficients depends on the
parity of $N$.

When $N=2,4,6,\dots$ is even then \be u_n=\left\{ { 4n(\alpha-n)
\quad \mbox{if} \quad n \quad \mbox{even}   \atop
4(N-n+1)(n+\beta-N-1) \quad \mbox{if} \quad n \quad \mbox{odd}}
\right . \lab{u-1_even} \ee and \be b_n=\left\{ {2N+1-\alpha-\beta
\quad \mbox{if} \quad n \quad \mbox{even}   \atop -2N-3
+\alpha+\beta \quad \mbox{if} \quad n \quad \mbox{odd}}  \right .
. \lab{b-1_even} \ee In compact form we have \be u_n = 4 [n]_{\xi}
[N-n+1]_{\eta}, \quad b_n= 2([n]_{\xi} + [N-n]_{\eta}) +
1-\alpha-\beta, \lab{rec_mu_form_e} \ee where
$$
\xi=\frac{\beta-N-1}{2}, \; \eta=\frac{\alpha-N-1}{2}
$$
and \be [n]_{\mu} = n+\mu(1-(-1)^n), \lab{mu_num} \ee are the
so-called "$\mu$-numbers" which appear naturally in problems
connected with the Dunkl operators \cite{Rosen}.

It is seen that $u_0=u_{N+1}=0$ as required for finite orthogonal
polynomials. The positivity condition $u_n>0, \; n=1,2,\dots, N$
is equivalent to the conditions \be \alpha>N, \; \beta>N.
\lab{pos_even_cond} \ee

Let us define the Bannai-Ito (BI) grid \be y_s=\left\{ {
-\alpha-\beta+2s+1 \quad \mbox{if} \quad s \quad \mbox{even},
\atop \alpha+\beta-2s-1 \quad \mbox{if} \quad s \quad \mbox{odd}}
\right . . \lab{even_grid_y} \ee The polynomials $R_n(x)$ are
orthogonal on the $N+1$ points $y_0, y_1, \dots, y_N$ of the BI
grid \be \sum_{s=0}^N w_s R_{n}(y_s) R_{m}(y_s) = \kappa_0 u_1 u_2
\dots u_n \: \delta_{nm}, \lab{ort_N_even} \ee where the discrete
weights are defined as \be w_{2s} = (-1)^s \frac{(-N/2)_s}{s!} \:
\frac{(1-\alpha/2)_s(1-\alpha/2-\beta/2)_s}{(1-\beta/2)_s
(N/2+1-\alpha/2-\beta/2)_s} \lab{w_N_e_s_e} \ee and \be w_{2s+1} =
(-1)^s \frac{(-N/2)_{s+1}}{s!} \:
\frac{(1-\alpha/2)_s(1-\alpha/2-\beta/2)_s}{(1-\beta/2)_s
(N/2+1-\alpha/2-\beta/2)_{s+1}}. \lab{w_N_e_s_o} \ee The
normalization coefficient is \be \kappa_0 = \frac{\left(
1-\frac{\alpha+\beta}{2} \right)_{N/2 }}{\left( 1-\frac{\beta}{2}
\right)_{N/2 }}. \lab{kap_even} \ee Assume that
$\alpha=N+\epsilon_1, \; \beta=N+\epsilon_2$, where
$\epsilon_{1,2}$ are arbitrary positive parameters. (This
parametrization corresponds to the positivity condition for the
dual -1 Hahn polynomials.) Then it is easily verified that all the
weights are positive $w_s>0, \: s=0,1,\dots,N$.

Moreover, the spectral points $y_s$ are divided into two
non-overlapping discrete sets of the real line: $$\{1-\delta,
-3-\delta, -7-\delta, \dots, -2N+1-\delta \}$$ and
$$
\{1+\delta, 5+\delta, 9+\delta, \dots, 2N-3+\delta \},
$$
where $\delta=\epsilon_1+\epsilon_2>0$. The first set corresponds
to $y_s$ with even $s$ and contains $1+N/2$ points; the second set
corresponds to $y_s$ with odd $s$ and contains $N/2$ points.

When $N=1,3,5,\dots$ is odd we have the BI grid $y_s$ \be
y_s=\left\{ { \alpha+\beta+2s+1 \quad \mbox{if} \quad s \quad
\mbox{even} \atop -\alpha-\beta-2s-1 \quad \mbox{if} \quad s \quad
\mbox{odd}} \right . \lab{odd_grid_y} \ee and the recurrence
coefficients \be u_n=\left\{ { 4n(N+1-n) \quad \mbox{if} \quad n
\quad \mbox{even}   \atop 4(\alpha+n)(\beta+N+1-n) \quad \mbox{if}
\quad n \quad \mbox{odd}} \right . \lab{u-1_odd} \ee and \be
b_n=\left\{ {-1-\alpha+\beta \quad \mbox{if} \quad n \quad
\mbox{even}   \atop -1 +\alpha-\beta \quad \mbox{if} \quad n \quad
\mbox{odd}}  \right . . \lab{b-1_odd} \ee In compact form we have
\be u_n = 4 [n]_{\xi} [N-n+1]_{\eta}, \quad b_n= 2([n]_{\xi} +
[N-n]_{\eta}) -2N-1-\alpha-\beta, \lab{rec_mu_form_o} \ee with
$\xi=\alpha/2, \: \eta=\beta/2$.

Note that the parameters $\xi$ and $\eta$ have different
expressions in \re{rec_mu_form_e} or \re{rec_mu_form_o}.

The positivity condition $u_n>0, \; n=1,2,\dots, N$ is equivalent
either to \be \alpha>-1, \; \beta>-1 \lab{pos_odd_cond} \ee or to
$\alpha<-N, \: \beta<-N$. It is sufficient to consider
\re{pos_odd_cond}.

The polynomials $R_n(x)$ are orthogonal on the set of $N+1$ points
$y_s$ \be \sum_{s=0}^N w_s R_{n}(y_s) R_{m}(y_s) = \kappa_0 u_1
u_2 \dots u_n \: \delta_{nm}, \lab{ort_N_odd} \ee where the
discrete weights are defined as \be w_{2s} = (-1)^s
\frac{(-(N-1)/2)_s}{s!} \:
\frac{(1/2+\alpha/2)_s(1+\alpha/2+\beta/2)_s}{(1/2+\beta/2)_s
(N/2+3/2+\alpha/2+\beta/2)_s} \lab{w_O_e_s_e} \ee and \be w_{2s+1}
= (-1)^s \frac{(-(N-1)/2)_s}{s!} \:
\frac{(1/2+\alpha/2)_{s+1}(1+\alpha/2+\beta/2)_s}{(1/2+\beta/2)_{s+1}
(N/2+3/2+\alpha/2+\beta/2)_s}. \lab{w_O_e_s_o} \ee The
normalization coefficient is \be \kappa_0 = \frac{\left(
1+\frac{\alpha+\beta}{2} \right)_{(N+1)/2 }}{\left(
\frac{\beta+1}{2} \right)_{(N+1)/2 }}. \lab{kap_odd} \ee

Assume that $\alpha=-1+\epsilon_1, \; \beta=-1+\epsilon_2$, where
$\epsilon_{1,2}$ are arbitrary positive parameters. (This
parametrization corresponds to the positivity condition for the
dual -1 Hahn polynomials for $N$ odd.) Then it is easily verified
that the weights are positive $w_s>0, \: s=0,1,\dots,N$.

Again, the spectral points $y_s$ are divided into two
non-overlapping discrete sets of the real line: $$\{-1-\delta,
-5-\delta, -9-\delta, \dots, -2N+1-\delta \}$$ and
$$
\{-1+\delta, 3+\delta, 7+\delta, \dots, 2N-3+\delta \},
$$
where $\delta=\epsilon_1+\epsilon_2>0$. Both sets contain
$(N-1)/2$ points.

The dual -1 Hahn polynomials can be explicitly expressed in terms
of the ordinary dual  Hahn polynomials; depending on whether $N$
and $n$ are even or odd, the formulas take different forms (see
\cite{-1_Hahn} for details).

In \cite{TVZ_para} it was shown that the -1 dual Hahn polynomials
appear as Clebsch-Gordan coefficients for the irreducible
representations of the algebra $sl_{-1}(2)$  which is the $q=-1$
limit of the quantum algebra $sl_q(2)$. The algebra $sl_{-1}(2)$
is generated by 4 operators $J_0, J_{+}, J_{-},R$ satisfying the
commutation relations \be [J_0,J_{\pm}]= \pm J_{\pm}, \quad [R,
J_0]=0, \quad \{J_{+}, J_{-} \}=2 J_0, \quad \{R , J_{\pm}\} =0,
\lab{comm_J} \ee where $[A,B]=AB-BA$ and $\{A,B\}=AB+BA$. The
operator $R$ is an involution operator, i.e. it satisfies the
property \be R^2=I . \lab{inv_R} \ee The Casimir operator $Q$,
commuting with $J_0$ and $J_{\pm}$ is \cite{TVZ_para} \be
Q=J_{+}J_{-} q^{-J_0} -\frac{2}{(q^2-1)(q-1)}
(q^{J_0-1}+q^{-J_0}). \lab{Q_q} \ee The algebra $sl_{-1}(2)$
admits a nontrivial addition rule (i.e. coproduct) \be \t
J_0=J_0\otimes {I} + {I} \otimes J_0, \quad \t J_{\pm } =
J_{\pm}\otimes {R} + {I} \otimes J_{\pm}, \quad \t R = {R} \otimes
{R}. \lab{add_rule} \ee such that the operators $\t J_0, \t
J_{\pm}, \t R$ again satisfy relations \re{comm_J}.

The Clebsch-Gordan coefficients (CGC) of $sl_{-1}(2)$ arise as
overlap coefficients between two canonical bases  in the coproduct
representation space. For details on the identification of the CGC
with -1 dual Hahn polynomials see \cite{TVZ_para}.

\section{Perfect state transfer and orthogonal polynomials}
\setcounter{equation}{0} Consider the spin chain with Hamiltonian
\be H=\frac{1}{2} \: \sum_{l=0}^{N-1} J_{l+1}(\sigma_l^x
\sigma_{l+1}^x + \sigma_l^y \sigma_{l+1}^y) + \frac{1}{2} \:
\sum_{l=0}^N b_l(\sigma_l^z +1), \lab{H_def} \ee where $J_l>0$ are
the constants coupling the sites $l-1$ and $l$ and $b_l$ are the
strengths of the magnetic field at the sites $l$
($l=0,1,\dots,N$). The symbols $\sigma_l^x, \: \sigma_l^y,\:
\sigma_l^z$ stand for the Pauli matrices.

Introduce the basis vectors
$$
|e_n ) = (0,0,\dots, 1, \dots, 0), \quad n=0,1,2,\dots,N,
$$
where the only 1 (spin up) occupies the $n$-th position. In that
basis, the restriction $J$ of $H$ to the one-excitation subspace
is given by the following $(N+1) \times (N+1)$  Jacobi matrix

\[ J= \left( \begin{array}{ccccc}
b_{0} & J_1 & 0 & \dots  & 0 \\
J_{1} & b_{1} & J_2 & \dots  & 0 \\
\dots & \dots  & \dots & \dots &    \dots \\
 0 & 0 & \dots & J_N & b_N \end{array} \right)\]

Equivalently, we have. \be J |e_n) = J_{n+1} |e_{n+1}) + b_n |e_n)
+ J_{n} |e_{n-1}). \lab{Je} \ee The boundary conditions \be
J_0=J_{N+1}=0 \lab{J0} \ee are assumed.

Let $x_s,\; s=0,1,\dots, N$ denote the eigenvalues of the matrix
$J$. They are all real and nondegenerate. They are labeled in
increasing order, i.e. $x_0<x_1<x_2<\dots x_N$.

To the Jacobi matrix $J$ one can associate the monic orthogonal
polynomials $P_n(x)$ defined by the 3-term recurrence relation \ba
&& P_{n+1}(x) + b_n P_n(x) + u_n P_{n-1}(x) = xP_n(x), \nonumber
\\ && n=0,1, \dots, N, \quad P_{-1}=0, \; P_0(x)=1, \lab{recP} \ea
where $u_n = J_n^2>0$.

$P_{N+1}(x)$ is the characteristic polynomial \be P_{N+1}(x) =
(x-x_0)(x-x_1) \dots (x-x_N). \lab{P_N+1} \ee The orthogonality
relation reads \be \sum_{s=0}^N P_n(x_s) P_m(x_s) w_s = h_n \:
\delta_{nm}, \lab{ort_P} \ee where
$$
h_n = u_1 u_2 \dots u_n
$$
and the discrete weights $w_s>0$ are uniquely determined by the
recurrence coefficients $b_n, u_n$.

Perfect state transfer (PST) occurs \cite{Kay} if there exists a
time $T$ such that \be e^{iTJ} |e_0) = e^{i \phi} |e_N ),
\lab{pqc} \ee where $\phi$ is a real number. This means that the
initial state $|e_0 )$  evolves into the state $|e_N )$ (up to an
inessential phase factor $e^{i \phi}$).

It is known that the PST property is equivalent to the two
conditions \cite{Kay}:

(i) the eigenvalues $x_s$ satisfy  \be x_{s+1} - x_s=
\frac{\pi}{T} M_s, \lab{xxM} \ee where $M_s$ are positive odd
numbers.

(ii) the matrix $J$ is mirror-symmetric $RJR =J$, where the matrix $R$ (reflection matrix) is
\[ R= \left( \begin{array}{ccccc}
0 & 0 & \dots & 0  & 1 \\
0 & 0 & \dots & 1  & 0 \\
\dots & \dots  & \dots & \dots &    \dots \\
 1 & 0 & \dots & 0 & 0 \end{array} \right)\]

Property (ii) is equivalent to the property \cite{VZ_PST}

(ii') the weights $w_s$ (up to a normalization) are given by the
expression \be w_s = \frac{1}{|P_{N+1}'(x_s)|}>0 . \lab{w_s_kap}
\ee

\section{$XX$ spin chains with PST and and dual -1 Hahn OPs}
\setcounter{equation}{0} We now wish to determine the
circumstances for which the Jacobi matrix $J$ corresponding to the
-1 dual Hahn polynomials will possess the PST property.

Crucial is the mirror symmetry condition (ii). In terms of the
recurrence coefficients it means \be u_{N-n+1}=u_n, \quad
b_{N-n}=b_n \lab{mirror} \ee for all $n=0,1, \dots N$.

Consider first the case of odd $N$. In this case formulas
\re{rec_mu_form_o}  immediately imply that conditions \re{mirror}
hold iff $\xi=\eta$, or, equivalently, iff $\alpha=\beta$. Under
this condition we have \be u_n = 4 [n]_{\alpha/2}
[N-n+1]_{\alpha/2}, \quad b_n= -1. \lab{ub_odd} \ee The positivity
condition for the measure is $\alpha<-1$.

The spectrum $x_s$ of the Jacobi matrix $J$ coincides with the BI
grid \re{odd_grid_y} and consists of two uniform subgrids $G_{-}$
and $G_{+}$ (with a step of 4 between the neighbor points).

The subgrid $G_{-}$ consists of the $(N+1)/2$ points $x_0<x_1<
\dots <x_{(N-1)/2}$: \be G_{-}=\{-2N+1+\delta, -2N+5+\delta,
\dots, -1-\delta \}. \lab{G-} \ee The subgrid $G_{+}$ consists of
the $(N+1)/2$ points $x_{(N+1)/2}<x_{(N+3)/2}< \dots <x_{N}$: \be
G_{+}=\{-1+\delta, 4 +\delta, \dots, 2N-3+\delta,\} \lab{G+} \ee
where $\delta =2(\alpha+1)>0.$ There is a gap of length $2 \delta=
4(\alpha+1)>0$ between these 2 subgrids.

Condition (i) in this case is equivalent to the restriction
\be
\alpha= \frac{M_2}{M_1}, \quad M_2>M_1, \lab{res_al_od} \ee
where $M_2$ is even and $M_1$ is odd ($M_2$ and $M_1$ are assumed to be coprime).

The Jacobi matrix with coefficients \re{ub_odd} and with the PST
property, was first introduced  in \cite{Shi}. In \cite{SJ} the
corresponding orthogonal polynomials $P_n(x)$ were related to the
dual Hahn polynomials. We see, that in fact these polynomials
coincide with the dual -1 Hahn polynomials for $N$ odd. Note that
in \cite{Shi}, \cite{SJ} the diagonal coefficients vanish $b_n=0$.
This corresponds to a simple shift of the argument of the
polynomials $P_n(x) \to P_n(x-1)$ as is seen from \re{ub_odd}.

Consider now the case of $N$ even. The recurrence coefficients are given by formulas \re{rec_mu_form_o}

The mirror symmetry condition \re{mirror} is again equivalent to
the condition $\alpha=\beta$. We then  have \be u_n= 4 [n]_{\xi}
[N-n+1]_{\xi}, \quad b_n = 4 \xi (-1)^{n+1} -1, \lab{ub_even} \ee
where $2\xi=\alpha-N-1$. The positivity condition implies
$\xi>-1/2$.

The spectrum $x_s$ of the corresponding Jacobi matrix $J$
coincides with the BI grid \re{even_grid_y}  and consists of two
uniform grids (with a step of 4 between the neighbor points)
containing $N+1$ points and $N$ points. There is a gap of length
$4(\alpha-N)>0$ between these sets.

Condition (i) in this case is equivalent to the restriction
\be
\alpha=N + \frac{M_1}{M_2}, \lab{res_even} \ee
where $M_1,M_2$ are positive coprime odd integers.

This example seems to be have been overlooked. Note that in
contrast to the case with $N$ odd, the diagonal recurrence
coefficients $b_n$ in \re{ub_even} are not constant. The only case
when $b_n=-1$ occurs when $\xi=0$ which corresponds to the
Krawtchouk polynomials.

It is interesting to point out that there is a connection based on
the Christoffel transform between these two spin models with PST.
To see this, suppose we have two models $A$ and $B$.  Assume that
the spin chain $A$ with PST corresponds to the Jacobi matrix $J_A$
with $N+1$ spectral points $x_0,x_1, \dots,x_N$ that satisfy
condition (i). Similarly, take spin chain $B$ with PST to
correspond to the Jacobi matrix $J_B$ with  $N$ eigenvalues $x_0,
x_1,x_2, \dots, x_{N-1}$ also satisfying (i). The matrix $J_A$ has
dimension $N+1 \times N+1$ while the matrix $J_B$ has dimension $N
\times N$ and the spectrum of $J_B$ differs from the spectrum of
$J_A$ by  the elimination of the level $x_N$. It is then easy to
show \cite{VZ_PST} that the matrix $J_B$ is obtained from the
matrix $J_A$ by a Christoffel transform. Equivalently, this means
that the monic orthogonal polynomials $\t P_n(x)$ corresponding to
the matrix $J_B$ are obtained from the polynomials $P_n(x)$
corresponding to matrix $J_A$ by the formula \be \t P_n(x) =
\frac{P_{n+1}(X) - K_n P_n(x)}{x-x_N}, \lab{CT} \ee where \be K_n
= \frac{P_{n+1}(x_N)}{P_n(x_N)}. \lab{K_def} \ee Formula \re{CT}
is equivalent to the well known Christoffel transform for the
orthogonal polynomials \cite{Sz}. The corresponding recurrence
coefficients can be obtained by the formulas \cite{VZ_PST} \be
\tilde u_n = u_n \frac{K_n}{K_{n-1}}, \quad \tilde b_n = b_{n+1}
+K_{n+1}-K_n . \lab{CT_UB} \ee

Returning to our systems, assume that the Jacobi matrix $J_A$ has
size $N+1 \times N+1$ with odd $N$ and corresponds to the spin
chain with coefficients \re{ub_odd}. The spectrum of this matrix
consists of $N+1$ points $x_s$ given by \re{G-}, \re{G+}. This
matrix corresponds to the PST model proposed in \cite{Shi}. The
Jacobi matrix $J_B$ has size $N \times N$ and its spectrum
consists of the two subgrids: \be \t G_{-} =\{x_0, x_1,\dots,
x_{(N-1)/2} \} = G_{-} \lab{G-B} \ee and \be \t G_{+} =
\{x_{(N+1)/2}, x_{(N+3)/2}, \dots x_{N-1} \}, \lab{G+B} \ee i.e.
the subgrid $\t G_{-}$ coincides with $G_{-}$ and contains
$(N+1)/2$ points. The subgrid $\t G_{+}$ contains $(N-1)/2$
points.

In order to obtain the recurrence coefficients $\t u_n, \t b_n$
corresponding to the matrix $J_B$  we need the expression for the
coefficients $K_n$ resulting from  \re{K_def}. A simple
calculation gives \be K_n=\left\{ {2(N+\alpha-n) \quad \mbox{if}
\quad n \quad \mbox{even} \atop 2(N-n) \quad \mbox{if} \quad n
\quad \mbox{odd}} \right . . \lab{K_n} \ee Equivalently, in terms
of $\mu$-numbers \re{mu_num}, we have \be K_n = 2[N-n]_{\alpha/2}.
\lab{K_mu} \ee Substituting this expression into formulas
\re{CT_UB} we obtain \be \t u_n = 4[n]_{\alpha/2}
[N-n]_{\alpha/2}, \quad \t b_n = -3 -2 (-1)^n \alpha . \lab{t_ub}
\ee These recurrence coefficients correspond to \re{ub_even}
(replacing $\alpha \to 2\xi$ and performing a shift of the
coefficient $b_n$).

We have thus demonstrated how the spin chain with the PST property
for even $N$ can be obtained from the corresponding spin chain for
odd $N$.

It was shown in \cite{VZ_PST}  that provided the eigenvalues $x_s,
\: s=0,1,\dots$ of the Jacobi matrix $J$ satisfy the condition
(i), the weight function $w_s$ constructed from \re{w_s_kap}
determines uniquely the mirror-symmetric matrix $J$ and hence, the
spin chain with PST.

Thus, the only spin chain with the PST property, corresponding to
the eigenvalues $x_s$ of the  BI type
\re{even_grid_y},\re{odd_grid_y} are those whose Jacobi matrices
$J$ generate the dual -1 Hahn polynomials with  $\beta=\alpha$. We
should stress, that the positivity conditions \re{pos_even_cond}
and \re{pos_odd_cond} are crucial. They imply, in particular that
the BI grid consists of two uniform subgrids separated by a
positive gap. Only for BI grids of such kind  can one associate
the dual -1 Hahn polynomials uniquely. More general BI grids with
an overlap of the two subgrids lead to more complicated orthogonal
polynomials. This problem will be considered elsewhere.

\bigskip\bigskip
{\Large\bf Acknowledgments}
\bigskip

AZ thanks Centre de Recherches Math\'ematiques (Universit\'e de
Montr\'eal) for hospitality. The authors would like to thank
Mathias Christandl for stimulating discussions. The research of LV
is supported in part by a research grant from the Natural Sciences
and Engineering Research Council (NSERC) of Canada.

\newpage

\bb{99}

\bi{Albanese} C.Albanese, M.Christandl, N.Datta, A.Ekert, {\it Mirror inversion of quantum states in linear registers}, Phys. Rev. Lett. {\bf 93} (2004), 230502.








\bi{Kay} A.Kay, {\it A Review of Perfect State Transfer and its Application as a Constructive Tool}, arXiv:0903.4274.

\bi{KS} R.Koekoek, R.Swarttouw, {\it The Askey-scheme of hypergeometric orthogonal polynomials and its q-analogue},
Report no. 98-17, Delft University of Technology, 1998.


\bi{Rosen} M. Rosenblum, {\it Generalized Hermite Polynomials and
the Bose-like Oscillator Calculus}, in: Oper. Theory Adv. Appl.,
vol. {\bf 73}, Birkhauser, Basel, 1994, pp. 369--396.
ArXiv:math/9307224.

\bi{Shi} T. Shi, Y.Li , A.Song, C.P.Sun, {\it Quantum-state transfer via the ferromagnetic chain in a spatially modulated
field}, Phys. Rev. {\bf A 71} (2005), 032309, 5 pages, quant-ph/0408152.

\bi{SJ} N.Stoilova, J.Van der Jeugt, {\it An exactly solvable spin chain related to Hahn polynomials}, SIGMA {\bf 7} (2011), 033, 13 pages.

\bibitem{Sz} G. Szeg\H{o}, Orthogonal Polynomials, fourth edition,  AMS, 1975.

\bi{BI} S.Tsujimoto, L.Vinet and A.Zhedanov, {\it Dunkl shift
operators and Bannai-Ito polynomials}, arXiv:1106.3512.

\bi{-1_Hahn} S.Tsujimoto, L.Vinet and A.Zhedanov, {\it Dual -1
Hahn polynomials: "classical" polynomials beyond the Leonard
duality}, arXiv:1108.0132.

\bi{TVZ_para} S.Tsujimoto, L.Vinet and A.Zhedanov, {\it From $sl_q(2)$ to a Parabosonic Hopf Algebra}, SIGMA {\bf 7} (2011), 093, 13 pages. arXiv:1108.1603.




\bi{VZ_PST} L.Vinet and A.Zhedanov, {\it How to construct spin
chains with perfect state transfer}, arXiv:1110.6474.

\eb

\end{document}